\documentclass[twocolumn,superscriptaddress,nofootinbib,amsmath,amssymb,amsfonts]{aastex631}
\usepackage{graphicx}%
\usepackage{dcolumn}%
\usepackage{enumitem}%
\usepackage{bm}%
\usepackage{hyperref}%
\hypersetup{
  colorlinks=true,
  linkcolor=blue, 
  citecolor=cyan, 
  urlcolor=cyan   
}
\usepackage{colortbl}
\usepackage{amsmath,amssymb,amsfonts}
\usepackage{times}

\newcommand{\cf}{cf.~}
\newcommand{\ie}{i.e.,~}
\newcommand{\eg}{e.g.,~}

\begin{document}

\title{On the maximum mass and oblateness of rotating neutron stars with
  generic equations of state}

\author[0000-0002-9955-3451]{Carlo Musolino}
\affiliation{Institut f\"ur Theoretische Physik, Goethe Universit\"at, Max-von-Laue-Str. 1, 60438 Frankfurt am Main, Germany}
\author[0000-0002-8669-4300]{Christian Ecker}
\affiliation{Institut f\"ur Theoretische Physik, Goethe Universit\"at, Max-von-Laue-Str. 1, 60438 Frankfurt am Main, Germany}
\author[0000-0002-1330-7103]{Luciano Rezzolla}
\affiliation{Institut f\"ur Theoretische Physik, Goethe Universit\"at, Max-von-Laue-Str. 1, 60438 Frankfurt am Main, Germany}
\affiliation{School of Mathematics, Trinity College, Dublin 2, Ireland}
\affiliation{Frankfurt Institute for Advanced Studies, Ruth-Moufang-Str. 1, 60438 Frankfurt am Main, Germany}

\date{\today}

\begin{abstract}
A considerable effort has been dedicated recently to the construction of
generic equations of state (EOSs) for matter in neutron stars. The
advantage of these approaches is that they can provide model-independent
information on the interior structure and global properties of neutron
stars. Making use of more than $10^6$ generic EOSs, we asses the validity
of quasi-universal relations of neutron star properties for a broad range
of rotation rates, from slow-rotation up to the mass-shedding limit. In
this way, we are able to determine with unprecedented accuracy the
quasi-universal maximum-mass ratio between rotating and nonrotating stars
and reveal the existence of a new relation for the surface oblateness,
\ie the ratio between the polar and equatorial proper radii. We discuss
the impact that our findings have on the imminent detection of new binary
neutron-star mergers and how they can be used to set new and more
stringent limits on the maximum mass of nonrotating neutron stars, as
well as to improve the modelling of the X-ray emission from the surface
of rotating stars.
\end{abstract}



\section{Introduction}

Matter inside neutron stars is compressed by gravity to densities a few
times larger than the saturation density of atomic nuclei,
$n_s=0.16\,\mathrm{fm}^{-3}$, making them the most compact material
objects known in our present universe. In principle, the properties of
neutron stars are fully determined by the equation of state (EOS) of
dense and cold, neutron-rich baryonic and possibly quark matter. In
practice, however, because the EOS is known only with large
uncertainties, our knowledge of even the most basic properties suffers from
serious limitations that are only mildly mediated by astronomical
observations and laboratory experiments. Among the various properties of
neutron stars for any given EOS, the maximum mass of rotating $M_{\rm
  max}$ and nonrotating configurations (TOV), $M_{_{\rm TOV}}$, bear
particular significance both in gravity and in nuclear physics.

Direct mass and radius measurements set strict lower limits on the
maximum mass of neutron stars $M_{_{\rm TOV}}\gtrsim
2\,M_\odot$~\citep{Fonseca:2021wxt} and constrain their radii to $\simeq
11$-$14\,$km~\citep{Miller:2021qha, Riley:2021pdl}. Additional upper
bounds on neutron-star radii and tidal deformabilities have been deduced
from the first direct gravitational-wave detection of a binary
neutron-star merger GW170817 by the LIGO and Virgo
collaborations~\citep{Abbott2018a}. From this event, upper bounds on the
maximum mass $M_{_{\rm TOV}}\lesssim 2.33\,M_\odot$ have been derived
from the associated gamma-ray burst GRB170817A~\citep{Margalit2017,
  Rezzolla2017, Ruiz2017, Shibata2019}.

One of the reasons why accurate theoretical predictions of neutron-star
properties are difficult is that reliable calculations of the EOS are
currently only available from Chiral Effective Field Theory at densities
$n_b \lesssim n_s$~\citep[see, \eg][]{Hebeler:2013nza, Gandolfi2019,
  Keller:2020qhx, Drischler:2020yad} and in the opposite limit from
pQCD~\citep[see, \eg][]{Freedman:1976ub, Vuorinen:2003fs, Gorda:2021kme,
  Gorda:2021znl} at densities $n_b \gtrsim 40~n_s$, that are much larger
than those reached inside neutron stars~\citep[see
  also][]{Komoltsev:2021jzg}. Between these limits, the only available
options are to either build models that reproduce the expected behaviour
of QCD~\citep[see, \eg][for some recent attempts]{Beloin2019,
  Bastian:2020unt, Traversi2020, Li2021, Demircik:2021zll,
  Ivanytskyi:2022wln}, or to construct agnostic (model independent)
parametrizations for the EOS and constrain them with astronomical or
multi-messenger measurements.

Our incomplete knowledge of the EOS is partially compensated by a number
of quasi-universal relations, \ie essentially EOS-independent, that have
been found among certain neutron-star quantities over the years, both in
terms of isolated rotating and nonrotating stars~\citep[see,
  \eg][]{Yagi2013a, Doneva2014a, Haskell2014, Chakrabarti2014,
  Pappas2014, Breu2016, Weih2017, Konstantinou:2022, Nath2023} and of the
gravitational-wave signal from binary systems~\citep[see,
  \eg][]{Bauswein2011, Read2013, Bernuzzi2014, Takami2015, Rezzolla2016,
  Most2018b, Bauswein2019, Weih:2019xvw, Gonzalez2022}
and~\citet{Yagi2017} for a review. Clearly, the robustness of these
quasi-universal relations depends on the number of EOSs that are employed
in determining the relations. So far, the identification and study of
quasi-universal relations was often limited by the relatively small
number of available models. At the same time, a growing number of recent
works have explored the possibility of building very large sets of
model-independent EOSs that satisfy all known theoretical and
observational constraints and cover the physically allowed space of
EOSs~\citep{Annala2019, Landry2018} and of related quantities, such as
the sound speed~\citep{Altiparmak:2022, Ecker:2022b, Gorda:2022}, or the
conformal anomaly~\citep{Marczenko:2022jhl, Annala2023,
  Brandes:2023hma}. These EOSs are built either using generic piecewise
polytropes \citep[see, \eg][]{Kurkela2014, Lattimer2014, Steiner2017,
  Most2018, Tews2018a, Tews2018}, a parameterization of the sound
speed~\citep[see, \eg][]{Annala2019, Altiparmak:2022, Ecker:2022b} or
non-parametric Gaussian process regression~\citep[see,
  \eg][]{Gorda:2022,Gorda:2023usm}.

The purpose of this work is to exploit such large ensembles of generic
EOSs to revisit the validity of quasi-universal relations of isolated
slowly and rapidly rotating stars, and to study how the results depend on
input from pQCD and on recently measured bounds.

\section{Quasi-universal relations}

Among the quasi-universal relations found for isolated stars, one is
particularly relevant for its impact on the stability of stellar models
and was first discussed by~\citet{Breu2016} (BR16 hereafter). More
specifically, using a relatively small number of 28 tabulated EOSs, BR16
pointed out that the critical mass $M_{\rm crit}$, that is, the mass of
uniformly rotating neutron stars on the turning-point line, can be
expressed in a quasi-universal relation through the specific angular
momentum $j_{\rm crit}$ and that at the mass-shedding limit $j_{\rm Kep}$
[$j_{\rm crit}$ and $j_{\rm Kep}$ are also indicated as $\chi_{\rm crit}$
  and $\chi_{\rm Kep}$, respectively~\citep{Most2020d}]
\begin{align}
  \frac{M_{\rm crit}}{M_{_{\rm TOV}}} = 1 + a_2 \left(
  \frac{j_{\rm{crit}}}{j_{\rm Kep}} \right)^2 +a_4 \left(
  \frac{j_{\rm{crit}}}{j_{\rm Kep}} \right)^4 \,,
\label{eq:M_crit}
\end{align}
where $a_2 = 0.132$, $a_4 = 0.071$~\citep{Breu2016}. A few remarks are
worth making. First, expression ~\eqref{eq:M_crit} provides the stellar
mass along the so-called ``turning-point'' line \citep{Friedman88}, that
is, the line in the $(M,n_c)$ plane along which $\left.\partial M /
\partial n_c \right|_J = 0$, where $n_c$ is the central number
density. Second, because the turning-point criterion is a sufficient
criterion for dynamical instability, the importance of~\eqref{eq:M_crit}
is that it allows one to determine for any rotation rate the critical
mass above which a dynamical instability would trigger the collapse to a
rotating black hole. Third, when considering the maximum value allowed
for the specific angular momentum, Eq.~\eqref{eq:M_crit} marks the
maximum mass of stable uniformly rotating configurations, namely, $M_{\rm
  max}:=M_{\rm crit}(j_\mathrm{crit} = j_{\rm{Kep}})$. Finally, because
the turning-point criterion is only a sufficient but not a necessary
criterion for dynamical instability, the maximum mass at the upper end of
the dynamical-instability line is actually set by the upper end of the
so-called ``neutral-stability'' line. Such a mass is normally slightly
larger than $M_{\rm max}$ and is attained at somewhat smaller central
densities~\citep[see][for a discussion]{Takami:2011, Weih2017}.

\begin{figure*}
  \center
  \includegraphics[height=.35\textwidth]{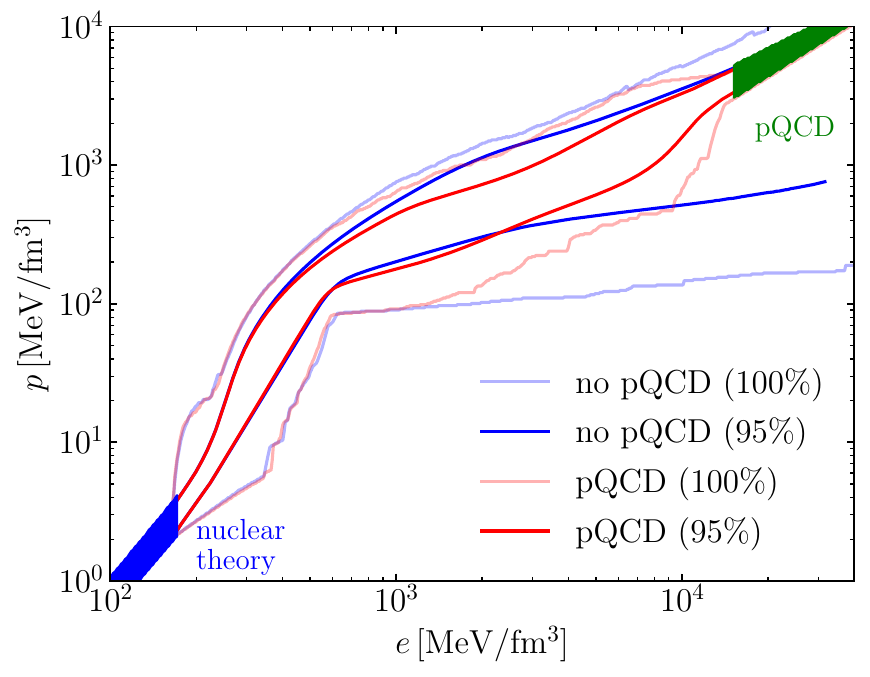}\quad\quad
  \includegraphics[height=.35\textwidth]{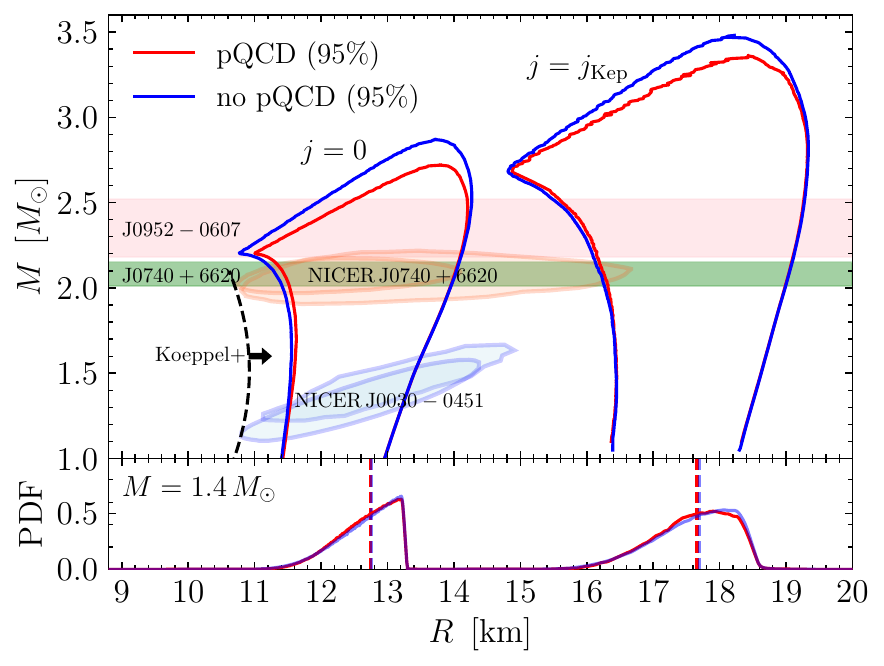}
  \caption{\textit{Left:} $95\%$ (dark colors) and $100\%$ (light colors)
    confidence-level contours for an ensemble of $10^6$ EOSs with
    $M^{(-)}_{_{\rm TOV}}=2.2~M_\odot$ for which the pQCD constraints are
    imposed (red) or not (blue). \textit{Right:} Corresponding $95\%$
    confidence-level contours for the mass-radius relations of
    nonrotating stars ($j=0$) and stars rotating at the mass-shedding
    frequency ($j=j_{\rm Kep}$; in this case the equatorial proper radii
    are used). Shaded areas indicate the imposed astrophysical
    constraints or the limits from the threshold mass to
    collapse~\cite{Koeppel2019}; the bottom panels report slices of the
    PDFs at $M=1.4~M_\odot$. }
  \label{fig:MR}
\end{figure*} 

In practice, the quasi-universal relation \eqref{eq:M_crit} is most often
used to relate the endpoints of the critical line via the ratio of
the maximum masses of rotating and nonrotating configurations, namely,
\begin{equation}
  \label{eq:R_BR16}
  \mathcal{R}:=\frac{M_{\rm max}}{M_{_{\rm TOV}}} \,.
\end{equation}
Using Eq.~\eqref{eq:M_crit}, BR16 estimated the mass ratio to be
$\mathcal{R}=1.203\pm0.022$, in rough agreement with cruder estimates
obtained previously with a much smaller number of EOSs~\citep{Cook94b,
  Cook94c, Lasota1996}; similar values have later been reported by
\citet{Bozzola2019} ($1.15-1.31$), \citet{Demircik:2020jkc}
($1.227^{+0.031}_{-0.016}$) for EOSs with a phase transition, and
by~\citet{Li2023} for EOSs with heavy baryons ($1.20815$). In what
follows, we assess the validity of Eqs.~\eqref{eq:M_crit}
and~\eqref{eq:R_BR16} and refine the quasi-universal behaviour exploiting
a set of more than $10^6$ different EOSs\footnote{Interestingly, already
$\mathcal{O}(10^3)$ EOSs are sufficient to obtain results very similar to
those from the full set of $10^6$ EOSs. This is because the probability
density functions (PDFs) converge very rapidly, as demonstrated in the
Appendix.} that covers the entire physically allowed space of EOSs
consistent with constraints from dense-matter theory and neutron-star
observations.

For the construction of our EOSs we follow the procedure presented in a
number of previous works where the interested reader can find additional
details~\citep[]{Altiparmak:2022, Ecker:2022b, Ecker:2022}. What is
relevant to recall here is that we construct the EOSs with a 10-segment
parameterization of the sound speed in which we can either impose or not
the pQCD constraints at $\sim 40\,n_s$. From the astrophysics side, we
impose constraints coming from radius measurements by the NICER
collaboration on J0740+6620~\citep{Miller2021, Riley2021} and
J0030+0451~\citep{Miller:2019cac, Riley2019} by rejecting EOSs for which
$R_{2.0} < 10.75~\mathrm{km}$ or $R_{1.1}<10.8~\mathrm{km}$, where the
subscript indicates the corresponding gravitational mass of nonrotating
stars. In addition, we impose an upper bound on the binary tidal
deformability $\tilde{\Lambda}$ as deduced from the LIGO/Virgo detection
of GW170817 by rejecting all EOSs with $\tilde{\Lambda} > 720$ (low-spin
prior)~\citep{Abbott2018a} at a chirp mass $\mathcal{M}_{\rm chirp} =
1.186~M_\odot$ for mass ratios $q>0.73$. The constraints from these
astrophysical observations are imposed as sharp cutoffs on the EOS
ensemble, which, as shown by~\citet{Jiang2022}, leads to good agreement
of the important central part of the PDF obtained from a more expensive
Bayesian analysis. Finally, we impose a lower bound $M^{(-)}_{_{\rm
    TOV}}$ on the maximum mass of nonrotating neutron stars ($M_{_{\rm
    TOV}}$) by rejecting all models with $M_{_{\rm TOV}}$ below a
prescribed cutoff. Since this constraint has been shown to have a
significant impact on the space of allowed EOSs~\citep{Ecker:2022b}, we
have explored different values for $M^{(-)}_{_{\rm TOV}}$ to account for
the uncertainties on this bound.

Once the EOS ensemble has been generated, the corresponding models for
rotating stars are constructed with the RNS
code~\citep{Stergioulas95}. In particular, we solve for stationary and
axisymmetric equilibrium solutions of uniformly rotating perfect fluids
with varying central densities along constant angular-momentum sequences
between $J=0$ and $J=J_{\rm Kep}$, where $J_{\rm Kep}$ is the Keplerian
(or ``mass-shedding'') angular momentum and is a function of
$n_c$~\citep[see also][for similar analyses with a much smaller set of
  EOSs]{Konstantinou:2022, Jia2023}. The set of stellar models having
$J=J_{\rm Kep}$ is also referred to as the Keplerian limit because along
this sequence the angular velocity is the largest possible; any increase
of the angular momentum at constant central density would lead to a
shedding of mass at the equator. The endpoint of the Keplerian sequence
marks the maximum mass at a Keplerian frequency, $M_{\rm max, Kep}$ and
this is close to, but systematically larger than $M_{\rm max}$. The
relative difference between the two masses is rather small, \ie $M_{\rm
  max, Kep}/M_{\rm max} - 1 \lesssim 10^{-2}$, and therefore often
ignored~\cite[\eg][]{Annala:2022}, but it is conceptually important that
the two masses are kept distinct, as it is done here.

The left panel of Fig.~\ref{fig:MR} reports the $95\%$ (dark lines) and
$100\%$ (light lines) confidence-level contours for our ensemble of
$10^6$ EOSs with $M^{(-)}_{_{\rm TOV}}=2.2~M_\odot$ when the pQCD
constraints are imposed (red) or not (blue). The right panel, on the
other hand, shows the corresponding $95\%$ confidence-level contours for
the mass-radius relations of nonrotating stars ($j=0$) and stars rotating
at the mass-shedding limit ($j=j_{\rm Kep}$); in the rotating case, the
equatorial radii are employed (see Appendix).

\section{Results}
\label{sec:results}

\begin{figure}
 \includegraphics[width=.45\textwidth]{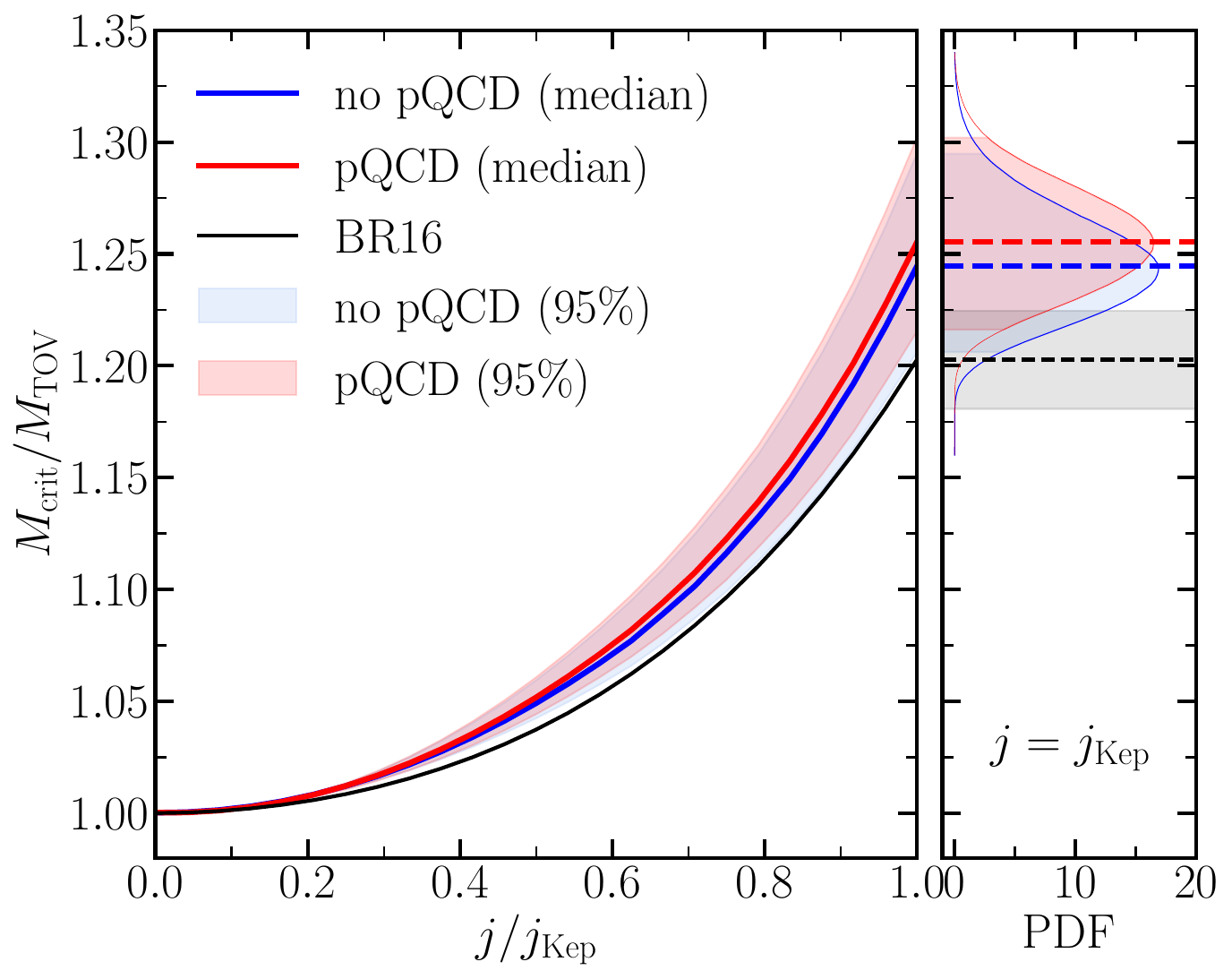}
 \caption{\textit{Left:} Quasi-universal relation of the gravitational
   mass along the critical line between. Red-shaded (blue-shaded) areas
   indicate the $95\%$ confidence intervals with (without) the pQCD
   constraint, while the solid lines mark the corresponding median
   values. The black solid line reports the prediction by~\cite{Breu2016}
   (BR16). \textit{Right:} One-dimensional slice of the PDFs for
   $j=j_{\rm Kep}$.}
  \label{fig:BR_Mtov2p2}
\end{figure}

Making use of our ensemble of $10^6$ EOSs, we have constructed more than
$10^8$ nonrotating and rotating stellar models up to the mass-shedding
limit. In this way, we have reconsidered several quasi-universal
relations governing the properties of neutron stars and first present the
outcome of this analysis for the mass-ratio
relation~\eqref{eq:M_crit}. In Fig.~\ref{fig:BR_Mtov2p2} we show results
for two different ensembles where the pQCD constraints are either imposed
(red colors) or not (blue colors). In both cases, we use a lower bound of
$M_{_{\rm TOV}}>2.2~M_{\odot}$ to ensure consistency with the
$1$-$\sigma$ confidence interval for the direct mass measurement of the
black-widow binary pulsar PSR~J0952-0607 
($2.35\pm0.17\,M_\odot$)~\citep{Romani:2022jhd}. More
specifically, the colored areas in the (large) left panel of
Fig.~\ref{fig:BR_Mtov2p2} show the ($95\%$) confidence intervals of the
PDF with (red) and without (blue) imposing the pQCD constraint, while the
solid lines of corresponding color mark the medians of the
distributions. The median values and the corresponding confidence
intervals of the distributions are still well approximated by
expression~\eqref{eq:M_crit}, where we use $a_2$ and $\mathcal{R}$ as
fitting parameters, thus fixing $a_4=\mathcal{R}-(a_2+1)$; the best-fit
coefficients are listed in Tab.~\ref{tab:results} of the
Appendix. Interestingly, the quadratic term in \eqref{eq:M_crit} provides
a very good approximation up to $j \sim 0.7\,j_{\rm Kep}$ and the quartic
contribution becomes essential only for larger values of $j$.

Overall, we find that the new and much larger set of EOSs shows again a
quasi-universal behaviour similar to that proposed by BR16, reported with
a black solid line in Fig.~\ref{fig:BR_Mtov2p2}. The variance of the new
universal relations is slightly larger than what was found by BR16, but
this is expected due to the much larger set of EOSs considered here. Note
also that the difference between the PDFs obtained when imposing and not
imposing the pQCD constraints is small but significant and has a clear
physical origin, which we discuss below. More importantly, the new
quasi-universal relations are slightly but systematically larger than
those reported in BR16 and, indeed, the median of the latter falls
outside the $95\%$ confidence interval for both the ensembles with and
without pQCD constraints. This is best seen in the right panel of
Fig.~\ref{fig:BR_Mtov2p2}, which reports the one-dimensional PDFs of
${M_{\rm crit}}/{M_{_{\rm TOV}}}$ at $j=j_{\rm Kep}$, \ie the PDFs of
$\mathcal{R}$. The medians of these PDFs are marked with red/blue dashed
lines for EOSs built with/without the pQCD constraints, and while they
differ by only $\simeq 1\%$, they are clearly separated. The reason for
this is that the pQCD constraint leads to a softening of the EOS at large
densities and, as a result, to smaller values of $M_{_{\rm TOV}}$.
Because the central density in maximally rotating stars is smaller than
in TOV stars, the impact on $M_{\rm max}$ is also smaller; the
combination of the two trends drives $\mathcal{R}$ towards slightly
larger values.  Finally, and more importantly, when considering the PDF
obtained with the pQCD constraint being imposed, we can extract a new
estimate for the mass ratio $\mathcal{R} = 1.255^{+0.047}_{-0.040}$,
which is therefore $\simeq 4\%$ larger than the BR16 estimate.

While the quasi-universality of the mass ratio $\mathcal{R}$ is robust,
it is clear that the constraints imposed on the EOS ensemble will
influence this ratio. To quantify this dependence, we also compute
$\mathcal{R}$ for a smaller ($M^{(-)}_{_{\rm TOV}}=2.0~M_\odot$) and a
larger ($M^{(-)}_{_{\rm TOV}}=2.35~M_\odot$) lower bound on $M_{_{\rm
    TOV}}$. The results are reported in
Tab.~\ref{tab:results}\footnote{Given the rapid convergence of the PDFs
with the number of EOSs (see Appendix), we have considered for these different
bounds $10^4$ EOSs.} and show that the median of the PDFs of
$\mathcal{R}$ increases monotonically with the imposed lower bound on
$M_{_{\rm TOV}}$. Stated differently, imposing a larger cutoff on the TOV
mass in the EOS ensemble drives the posterior for $M_{\rm max}$ upwards
faster than that of $M_{_{\rm TOV}}$. While this might seem
counter-intuitive at first, it is simple to understand. We recall that
imposing larger bounds on $M_{_{\rm TOV}}$ leads to a significant
stiffening of the EOSs, \ie in the energy-density range $0.5 <
e/(\mathrm{GeV\,fm^{-3}}) < 0.8$ [see Fig.~1 of \citet{Ecker:2022b}] and
that are relevant for the cores of rapidly rotating stars. At the same
time, such bounds have a negligible impact on energies $e\approx
1~\mathrm{GeV/fm^3}$, that are those relevant for the cores of
nonrotating stars~\citep{Altiparmak:2022}. This increase of $M_{\rm max}$
not balanced by an equal increase in $M_{_{\rm TOV}}$ leads to the
measured growth of $\mathcal{R}$ with $M^{(-)}_{_{\rm TOV}}$.

The new and more accurate estimate of $\mathcal{R}$ provides two
important tools to be employed when a binary neutron-star merger is
detected together with its electromagnetic counterpart. First, as
discussed by~\citet{Rezzolla2018} in the case of GW170817, the detection
of a gamma-ray burst counterpart following a binary neutron-star merger
can be taken as an indirect evidence for the formation of a black hole
from the remnant and the emergence of a relativistic jet from the ejected
matter~\citep{Gill2019}. This implies that the mass of the merger
remnant, properly reduced by the ejected rest-mass and the mass lost in
gravitational waves, is very close to $M_{\rm max}$, setting an upper
limit for it. In turn, making use of $\mathcal{R}$, this can be used to
set an upper limit on $M_{_{\rm TOV}}$. Following this logic and
employing the iterative procedure to account for an adaptive adjustment
of the mass ratio (see Appendix for details), we obtain the following
upper limit for the maximum mass of a nonrotating neutron star
$M^{(+)}_{\rm TOV} = 2.24^{+ 0.07}_{-0.11}$. Clearly, the same method can
be employed to further refine the estimate of $M^{(+)}_{\rm TOV}$ as new
detections of merging binary-neutron star systems will be made.

A second consequence of the new estimate of $\mathcal{R}$ follows from
the discussion made in~\citet{Most2020d}. In particular, once a
gravitational-wave merger event with a significant mass difference such
as GW190814 is detected, the knowledge of the total mass of the system
and of the universal mass ratio $\mathcal{R}$ allows one to set
constraints on the spin of the secondary [see Fig.~1 of
  \citet{Most2020d}]. More specifically, assuming that the secondary of
GW190814 was a neutron star at the merger~\citep[see][for a different
  conclusion]{Nathanail2021}, we can employ the new value of
$\mathcal{R}$ to improve the constraints on the dimensionless spin of the
secondary in GW190814 to be $0.52 \lesssim \chi_2 \lesssim 0.72$. Also in
this case, the methodology discussed above can be applied to new
detections of binaries with small mass ratios.

\begin{figure}
  \includegraphics[width=0.95\columnwidth]{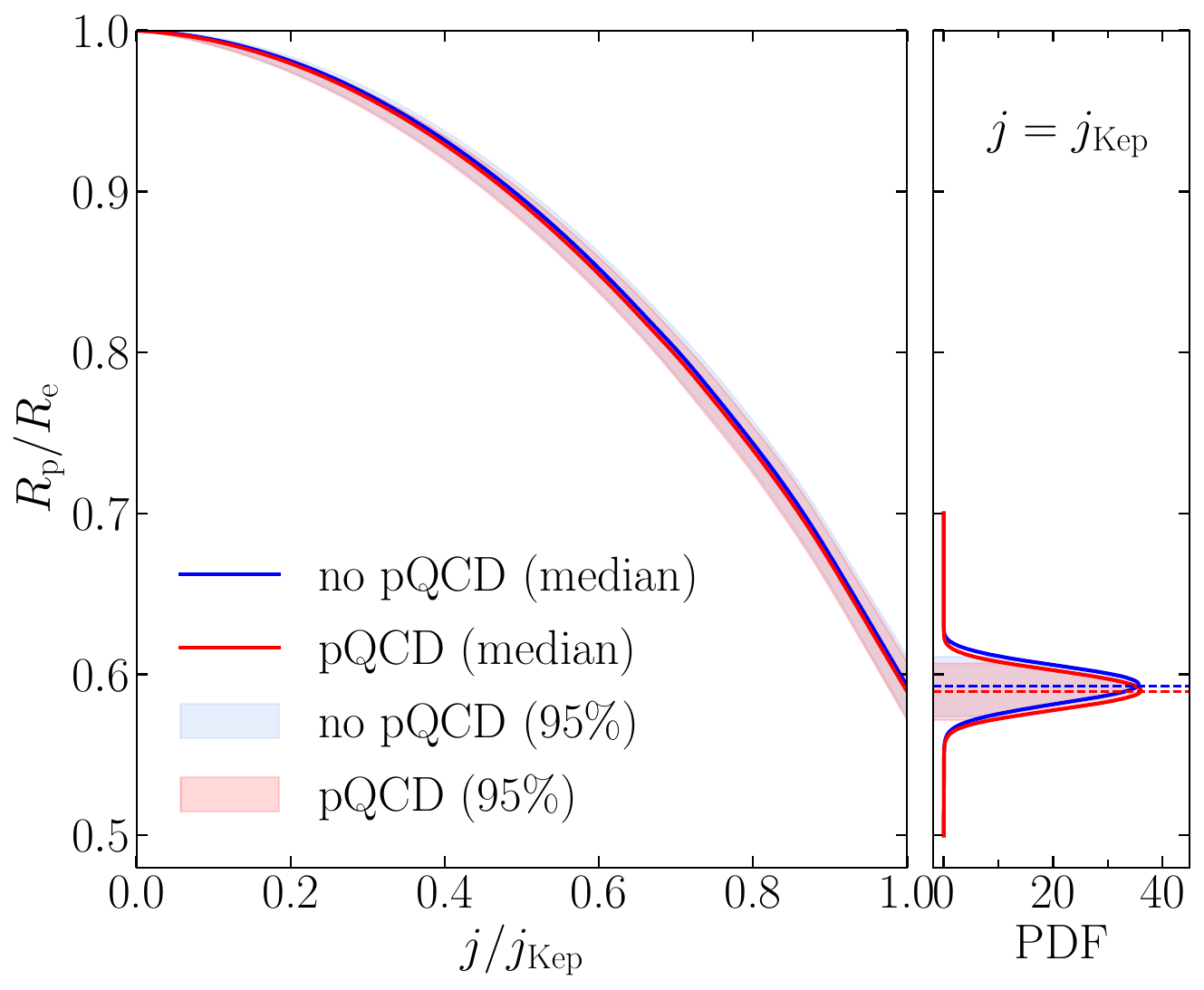}%
  \caption{The same as in Fig.~\ref{fig:BR_Mtov2p2} but for the novel
    quasi-universal relation for the ratio of the polar-to-equatorial
    radii.}
  \label{fig:Rp/Re}
\end{figure}

We next employ our large set of rotating-star models to assess the
validity of other quasi-universal relations and we start by reporting a
novel quasi-universal relation found for the surface oblateness, \ie the
ratio between the polar and equatorial proper radii, $R_{\rm p}/R_{\rm
  e}$. This ratio is obviously unity in the case of nonrotating stars and
decreases as the angular momentum is increased, since the equatorial
radius becomes larger and the quadrupolar deformation of the star
increases~\citep[see also][for different but equally interesting
  relations]{Konstantinou:2022}. The result of our analysis in this case
is presented in Fig.~\ref{fig:Rp/Re}, which is logically similar to
Fig.~\ref{fig:BR_Mtov2p2}, but now for the ratio $R_{\rm p}/R_{\rm
  e}$. Interestingly, the quasi-universal relation has an extremely small
variance and this is most probably due to the fact that $R_{\rm p}/R_{\rm
  e}$ depends most sensitively on the low-density part of the EOSs which
is rather well constrained. This hypothesis is also corroborated by the
fact that the result for this relation seems to be essentially
independent on whether or not the perturbative QCD constraint is imposed,
implying that the high-density portion of the EOS, where this constraint
has an impact, is not involved in the deformation seen in
Fig.~\ref{fig:Rp/Re}. Given the simple behavior of the relation, the
corresponding medians can be very accurately described by a second-order
polynomial of the form $R_p/R_e = 1 - b_2~(j/j_{\rm Kep})^2$, and the
corresponding best-fit coefficients are reported in
Tab.~\ref{tab:results} of the Appendix. The relevance of this result is
that it allows for a much more accurate modelling of the emission from
rotating neutron stars, such as those observed by NICER, where a precise
knowledge of the surface deformation is essential for the modelling of
the X-ray emission from the hot spots of rotating
stars~\citep{Morsink2007, Bogdanov2019b}. Furthermore, it can be employed
when studying black-widow systems, where rapid rotation is expected and
indeed observed [with a spin frequency of ${f=706\,\rm Hz}$,
  PSR~J0952-0607 is the second-fastest-spinning pulsar
  known~\citep{Romani:2022jhd} and has $j/j_{\rm Kep} \lesssim 0.46$].

Finally, we use our EOS ensemble to assess the validity of the so-called
``$I$-$Q$'' quasi-universal relations, where $I$ is the stellar moment of
inertia and $Q$ the mass-quadrupole moment. The IQ relations were first
investigated by~\citet{Yagi2013a} in the context of slowly rotating
neutron stars and subsequently extended to stars in rapid rotation
by~\citet{Doneva2014a, Chakrabarti2014}, and~\citet{Pappas2014}. The
latter, in particular, have found quasi-universal relations between $I$
and $Q$ also for rapidly rotating stars and expressed them in terms of
$j$. Since these functions were validated using a dozen of
nuclear-physics EOSs only, it is interesting and important to assess the
validity of such relations when employing a much larger set of
model-independent EOSs~\citep[see also][for similar recent
  work]{Nath2023}. Overall, we find that the fit proposed
by~\citet{Pappas2014} performs very well also with the much larger
ensemble of EOSs and the differences in the newly estimated best-fit
coefficients are of the order of $1\%$, with the largest differences of
$\sim 2.5\%$ being attained in the low-$Q$ and high-$j$ region of the
parameter space (see Appendix for details).

\section{Conclusion}
\label{sec:conclusions}

Motivated by recent advancements in model-agnostic sampling methods for
the EOS of neutron-star matter~\citep{Kurkela2014, Lattimer2014,
  Steiner2017, Most2018, Tews2018a, Annala2019, Altiparmak:2022}, we have
revisited the quasi-universal behaviour of isolated rotating and
nonrotating neutron stars by employing a very large sample of generic
EOSs constructed with a 10-segment parameterization of the sound
speed~\citep{Annala2019, Altiparmak:2022}. By exploring the whole
physically allowed space of solutions, our approach has the advantage of
being able to test rigorously whether a quasi-universality is present or
not, and to study the dependence of the quasi-universal behaviour on the
imposed constraints.

Our attention here has been focussed on the quasi-universal relation of
the mass along the critical line and, more in particular, on the mass
ratio $\mathcal{R}$ between the maximum masses of rotating and
nonrotating stars~\citep{Breu2016}. Our analysis has revealed that such a
relation is still valid when considering an ensemble of up to $10^6$ EOSs
obtained either imposing or not the constraints from pQCD at high
densities, and with different lower bounds the maximum mass of
nonrotating stars $M_{_{\rm TOV}}$. While the variance in the new
estimates is comparable but larger than in BR16, the median values are
a few percent larger than the original estimate of BR16. This is not
surprising given the comparatively small number of EOSs considered by
BR16. Furthermore, we have found that increasing the lower bound on
$M_{_{\rm TOV}}$ also leads to a larger value of $\mathcal{R}$, while
neglecting the pQCD constraints at high densities leads to a decrease of
this ratio.

Finally, we have employed our large set of EOSs to assess the validity of
known and novel quasi-universal relations. In particular, we have
revealed the existence of a new and tight quasi-universal relation for
the surface oblateness, which can be used to model more accurately the
emission from the hot spots on the surface of rotating neutron stars,
such as those observed by NICER. At the same time, we were able to
confirm the validity of the $I$-$Q$ relations over a broad range of
rotations, from slow rotation up to the mass-shedding limit, and to
improve the best-fit coefficients of the functional behaviour proposed
by~\citet{Pappas2014} in the low-$Q$ and high-$j$ region of the parameter
space.

Besides improving previously known results, the more accurate estimate of
$\mathcal{R}$ offers two important and direct applications whenever a
binary neutron-star merger is detected together with its electromagnetic
counterpart. In particular, it can be used to set new and tighter
constraints on the maximum mass of nonrotating stars and hence on the
EOS. For example, when considering GW170817, this implies a new upper
limit of $M^{(+)}_{\rm TOV} = 2.24^{+
  0.07}_{-0.11}~M_\odot$. Furthermore, in binaries with small mass ratio
it can be employed to set constraints on the dimensionless spin of the
secondary object when this is represented by a neutron star. For example,
when considering the case of GW190814, the new value of $\mathcal{R}$
implies a new bracketing interval of $0.52 \lesssim \chi_2 \lesssim
0.72$. The present data-taking runs of the LIGO-Virgo-Kagra collaboration
will hopefully provide us with a number of potential applications of
these findings.

\section*{Acknowledgements}
We are grateful to R. Duque for insightful discussions and comments. We also
thank R. Mallick for his careful reading of the manuscript.
Partial funding comes from the State of Hesse within the Research
Cluster ELEMENTS (Project ID 500/10.006), by the ERC Advanced Grant
``JETSET: Launching, propagation and emission of relativistic jets from
binary mergers and across mass scales'' (Grant No. 884631). CE
acknowledges support by the Deutsche Forschungsgemeinschaft (DFG,
German Research Foundation) through the CRC-TR 211 ``Strong-interaction
matter under extreme conditions''-- project number 315477589 -- TRR
211. LR acknowledges the Walter Greiner Gesellschaft zur F\"orderung
der physikalischen Grundlagenforschung e.V. through the Carl W. Fueck
Laureatus Chair. The calculations were performed on the local ITP
Supercomputing Clusters Iboga and Calea.

\section*{Data Availability}
Data is available upon reasonable request from the corresponding
author.


\section*{Appendix}
\label{sec:SM}

\subsection*{Overview of the stellar properties}
\label{app:star_prop}

\begin{figure*}
  \center
  \includegraphics[width=.32\textwidth]{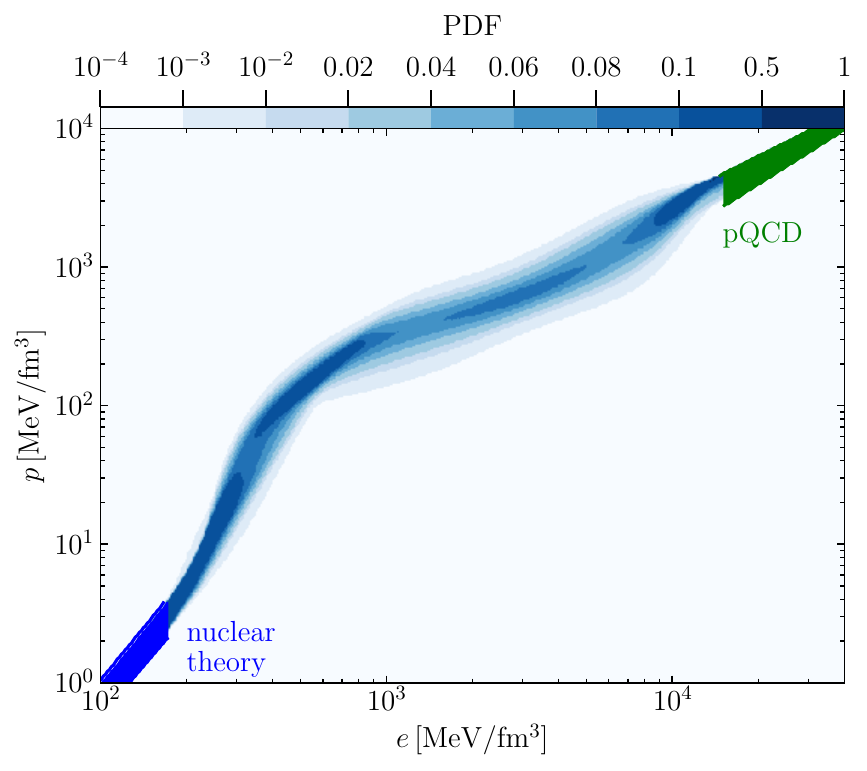}
  \includegraphics[width=.32\textwidth]{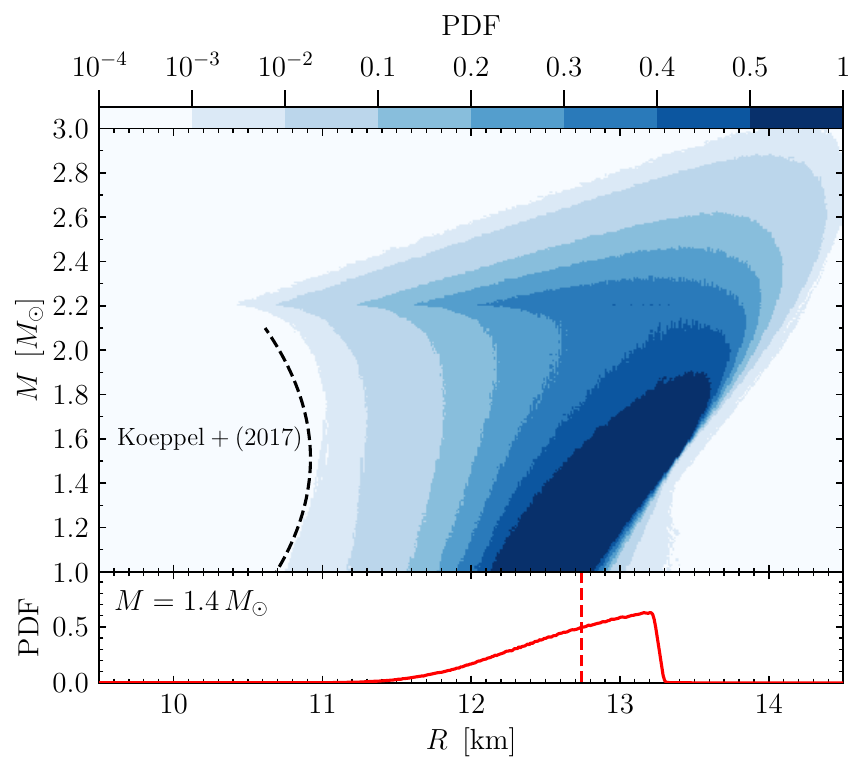}
  \includegraphics[width=.32\textwidth]{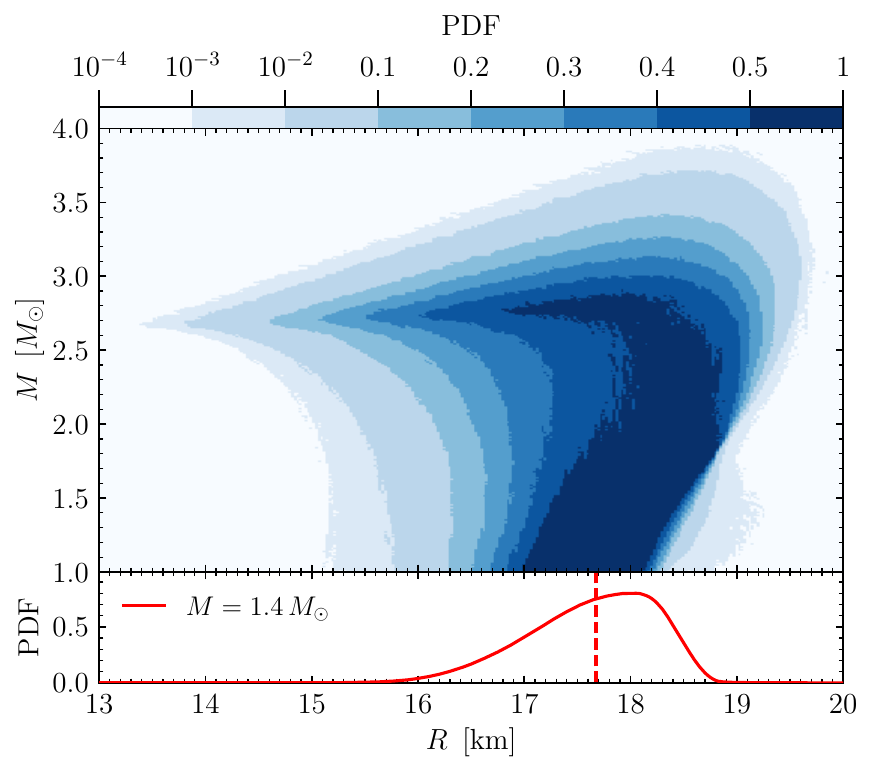}
  \includegraphics[width=.32\textwidth]{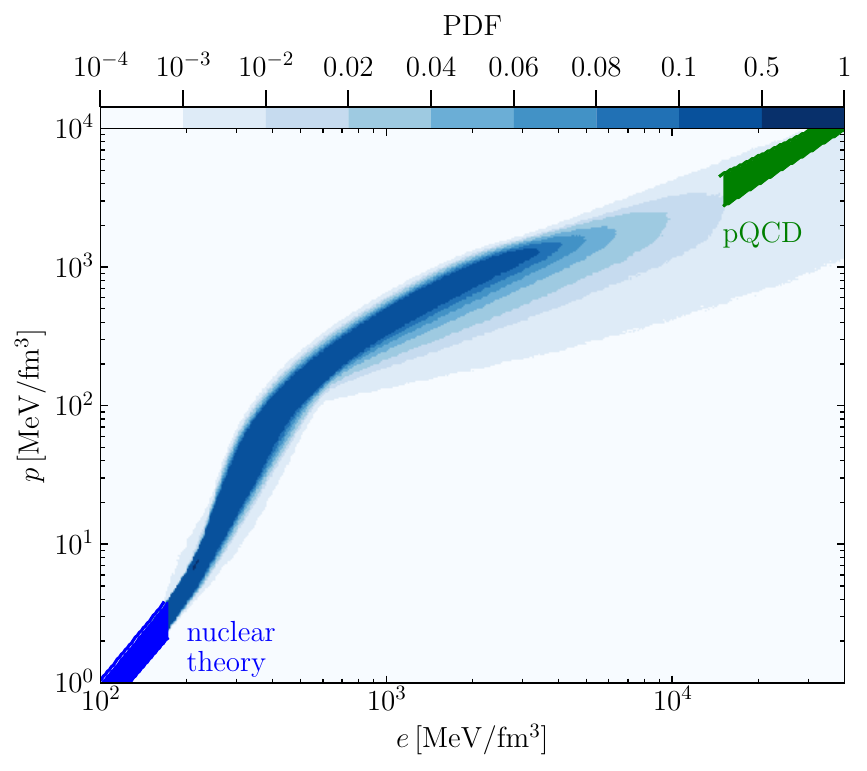}
  \includegraphics[width=.32\textwidth]{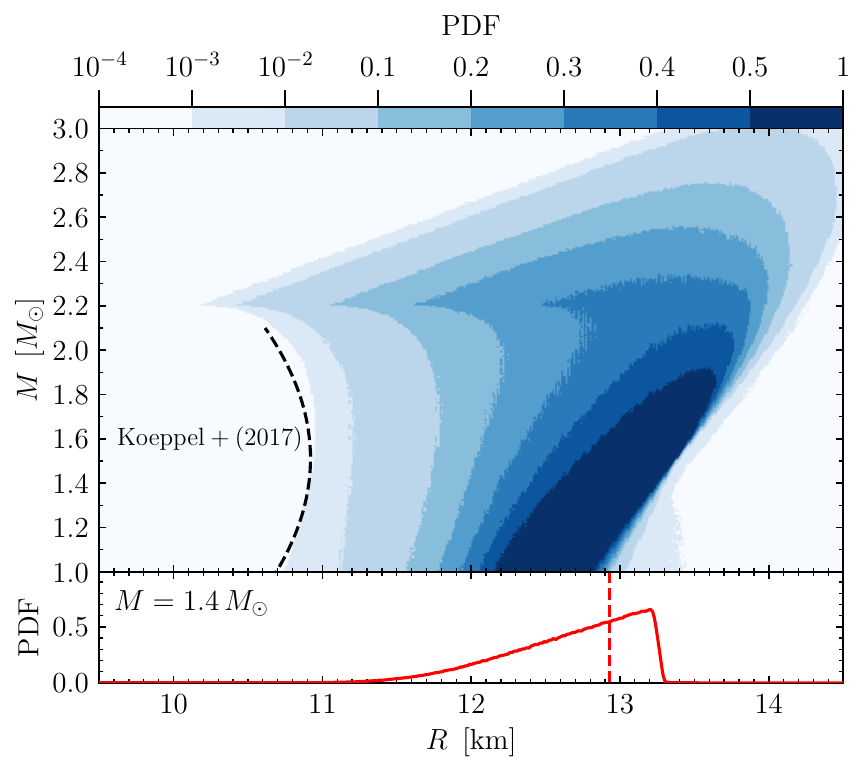}
  \includegraphics[width=.32\textwidth]{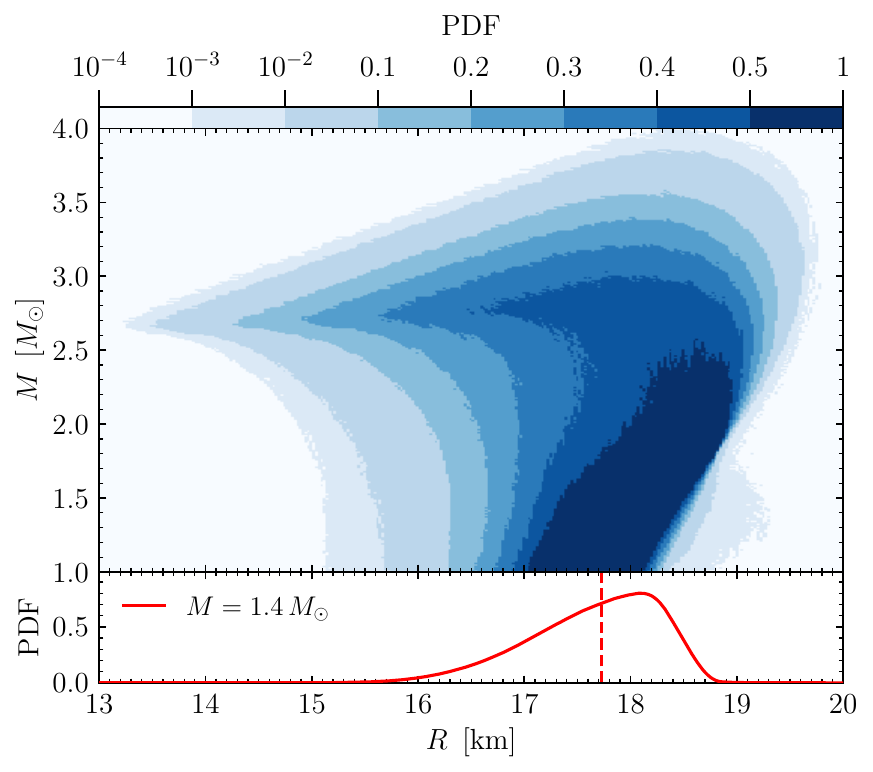}
  \caption{PDFs of the EOSs (left panel), of the mass-radius relation of
    nonrotating stars (middle panel) and of stars rotating at the
    mass-shedding frequency (right panel). The top and bottom row refer
    respectively to ensembles where the pQCD constraints are imposed or
    not; in all cases $M^{(-)}_{\rm TOV}=2.2~M_\odot$.}
  \label{fig:EOS_PDF}
\end{figure*}

In Fig.~\ref{fig:EOS_PDF} we show the PDFs for the EOSs (left panel) and
the corresponding mass-radius relation of static (middle panel) and
maximally rotating stars (right panel) when imposing (top row) or not
(bottom row) the pQCD constraint and assuming $M^{(-)}_{_{\rm
    TOV}}=2.2~M_\odot$. As clearly shown by the figure, the allowed
pressure band is influenced by imposing the pQCD constraint down to
energy densities $e\sim 800~\mathrm{MeV}$, which are relevant for the
cores of heavies nonrotating stars TOV and even for rapidly rotating
stars. In particular, the EOSs are allowed to be significantly stiffer
and with higher sound speeds [\ie with steeper curves in the $(p,e)$
  space] in the range $0.8-1.2~\mathrm{GeV}$ since the EOSs do not need
to be causally connected to the pQCD band at high densities. As can be
observed in the two rightmost columns of Fig.~\ref{fig:EOS_PDF}, and in
Fig.~\ref{fig:MR}, the effect of this stiffening in the intermediate to
high-density regime is to allow for higher masses and to increase the
radii of stars close to the stability limit. In particular, the lower
panels in these figures show that the distributions of radii for
$1.4\,M_\odot$ models varies only modestly whether or not the EOS
ensemble is restricted to be compatible with results of pQCD. On the
other hand, the contours of the distribution close to the critical limit
clearly show that the constraint has non-negligible impact on
TOVs~\citep[which was already shown in][]{Gorda:2022}, as well as on
rapidly rotating stars, which we show here for the first time.

\begin{figure*}
  \center
  \includegraphics[width=.9\textwidth]{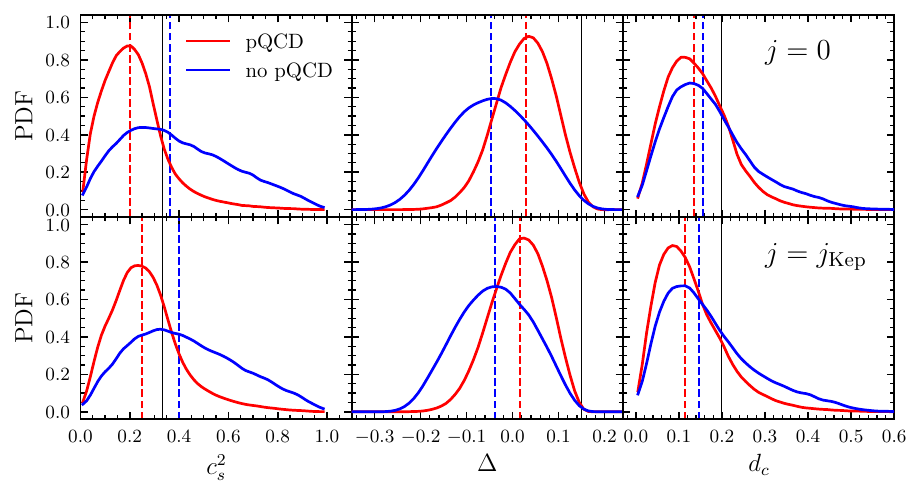}
  \caption{One-dimensional PDFs for the sound speed (left column), for
    the normalized trace anomaly (middle column) and for the measure of
    the non-conformality of the matter (right column). All quantities are
    computed at the center of $j=0$ (top row) and $j=j_{\rm Kep}$ (bottom
    row) stellar models. Vertical black lines are approximate upper
    bounds reported in~\citet{Annala2023} for of quark matter in pQCD at
    $n_b>40~n_s$.}
  \label{fig:1d_pdfs}
\end{figure*}

Similarly, in Fig.~\ref{fig:1d_pdfs} we report the one-dimensional PDFs
for the sound speed (left column), for the normalized trace anomaly
$\Delta=1/3-{p}/{e}$~\citep{Fujimoto:2022ohj_pub} (middle column) and for
the measure of the non-conformality of the matter $d_c := \sqrt{ \Delta^2
  + (\Delta^\prime)^2}$ (right column), where $\Delta^\prime$ is the
logarithmic derivative of $\Delta$ with respect to the energy density.
All quantities are computed at the center of $j=0$ (top row) and
$j=j_{\rm Kep}$ (bottom row) stellar models. The median values, together
with the $95\%$ error bars are reported in
Tab.~\ref{tab:results}. Overall, our results compare well with the
corresponding slices calculated by~\citet{Annala2023}, as well as with
the results of~\citet{Gorda:2022}, obtained with a Gaussian-process
regression technique~\citep{Landry2018}.

Finally, in Tab.~\ref{tab:results} we also report the median values with
95\% error bars for $\mathcal{R}$, the best-fit value for the coefficient
$a_2$, the central sound speed squared in either maximally rotating
rotating ($c_{s,{\rm c, Kep}}^2$) or static stars ($c_{s,{\rm
    c,{_{TOV}}}}^2$). The last four columns report the central values of
the conformal anomaly $\Delta_c$ and $d_c$ for various choices of
$M^{(-)}_{_{\rm TOV}}$, and with or without imposing the pQCD constraint.

\begin{table*}
  \setlength{\tabcolsep}{3pt} 
  \renewcommand{\arraystretch}{1.35} 
  \centering
  \footnotesize
 \begin{tabular}{cccccccccccc}
   \hline
   $M^{(-)}_{_{\rm TOV}}$ & pQCD & $\mathcal{R}$ &   $a_2$  & $b_2$ & $c_{s,{\rm c, Kep}}^2$       & $c_{s,{\rm  c,{_{TOV}}}}^2$      & $\Delta_{\rm c, Kep}$   & $\Delta_{\rm c,{_{TOV}}}$   & $d_{\rm c, Kep}$       & $d_{\rm c,{_{TOV}}}$      \\
   $[M_\odot]$ &  &  &  &  &  &  &  &  &  & \\
   \hline                                                                                                                                                                                                              
  $2.00$                   & yes & $1.248^{+0.054}_{-0.043}$ & $0.18^{+0.04}_{-0.02}$ & $0.41^{+0.02}_{-0.02}$    & $0.245^{+0.381}_{-0.195}$ & $0.208^{+0.369}_{-0.172}$ & $0.031^{+0.115}_{-0.136}$  & $0.038^{+0.117}_{-0.133}$  & $0.118^{+0.209}_{-0.095}$ & $0.130^{+0.180}_{-0.102}$ \\
			   & no  & $1.240^{+0.055}_{-0.043}$ & $0.16^{-0.04}_{-0.02}$ & $0.40^{+0.03}_{-0.02}$    & $0.377^{+0.491}_{-0.316}$ & $0.348^{+0.502}_{-0.306}$ & $-0.024^{+0.159}_{-0.162}$ & $-0.031^{+0.143}_{-0.174}$ & $0.146^{+0.286}_{-0.117}$ & $0.153^{+0.260}_{-0.124}$ \\
  $2.20$                   & yes & $1.255^{+0.047}_{-0.040}$ & $0.18^{+0.03}_{-0.02}$ & $0.41^{+0.02}_{-0.02}$    & $0.248^{+0.388}_{-0.195}$ & $0.201^{+0.360}_{-0.167}$ & $0.017^{+0.104}_{-0.129}$  & $0.029^{+0.108}_{-0.133}$  & $0.114^{+0.214}_{-0.094}$ & $0.135^{+0.180}_{-0.107}$ \\
			   & no  & $1.244^{+0.050}_{-0.039}$ & $0.17^{-0.04}_{-0.02}$ & $0.40^{+0.02}_{-0.02}$    & $0.398^{+0.470}_{-0.325}$ & $0.365^{+0.489}_{-0.314}$ & $-0.038^{+0.147}_{-0.153}$ & $-0.046^{+0.167}_{-0.166}$ & $0.146^{+0.288}_{-0.120}$ & $0.156^{+0.260}_{-0.126}$ \\
  $2.35$                   & yes & $1.260^{+0.041}_{-0.034}$ & $0.18^{-0.03}_{-0.02}$ & $0.41^{+0.02}_{-0.02}$    & $0.256^{+0.386}_{-0.198}$ & $0.195^{+0.326}_{-0.164}$ & $0.003^{+0.094}_{-0.121}$  & $0.019^{+0.100}_{-0.129}$  & $0.112^{+0.222}_{-0.093}$ & $0.138^{+0.180}_{-0.110}$ \\
			   & no  & $1.248^{+0.046}_{-0.034}$ & $0.16^{-0.04}_{-0.02}$ & $0.40^{+0.02}_{-0.01}$    & $0.419^{+0.458}_{-0.338}$ & $0.375^{+0.487}_{-0.322}$ & $-0.052^{+0.137}_{-0.142}$ & $-0.059^{+0.162}_{-0.158}$ & $0.152^{+0.285}_{-0.126}$ & $0.161^{+0.263}_{-0.131}$ \\
  \hline
 \end{tabular}
  \caption{Estimates for various properties of static and rotating
    neutron stars. The central values and the uncertainties correspond to
    the median and the $95\%$ confidence intervals of the PDF,
    respectively. Note that the upper and lower bounds of $a_2$ have been
    computed with~\eqref{eq:M_crit} using the corresponding upper and
    lower bounds of $\mathcal{R}$.}
  \label{tab:results}
\end{table*}

\subsection*{Convergence and an upper bound on $M_{_{\rm TOV}}$}
\label{app:convergence}

\begin{figure*}[htb]
  \center
  \includegraphics[width=0.5\textwidth]{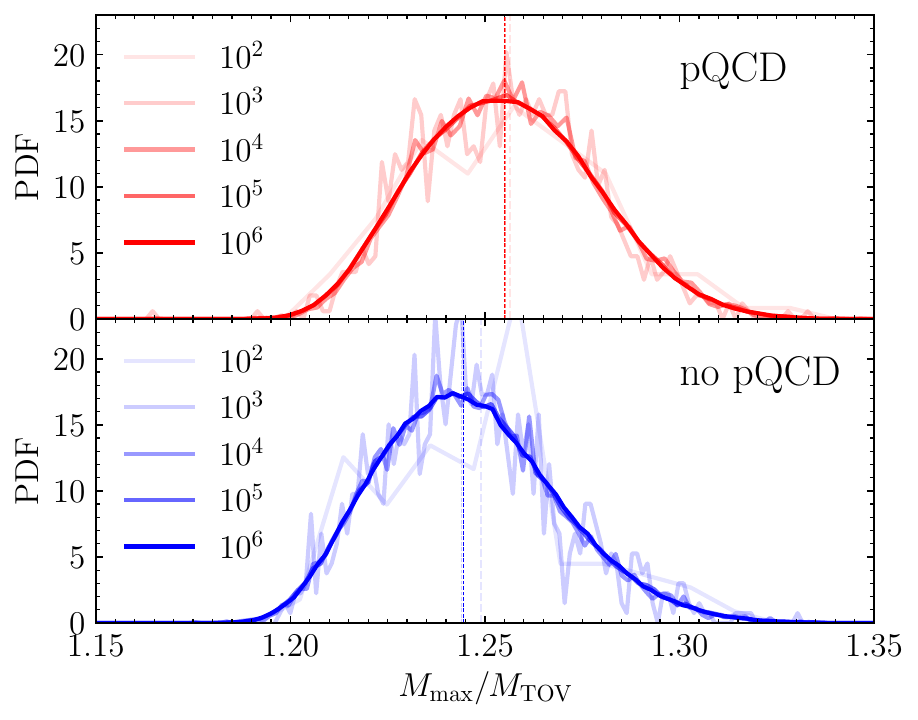}
  \caption{One-dimensional PDFs of $\mathcal{R}$ when varying the number
    of EOSs in the set. Note that $10^3$ EOSs are already sufficient to
    obtain a very small variance.}
  \label{fig:convergence}
\end{figure*}

To demonstrate the robustness and convergence of the results for
$\mathcal{R}$ obtained with our sampling procedure we compare in
Fig.~\ref{fig:convergence} the one-dimensional PDFs obtained with our
procedure for the ratio $\mathcal{R}$; obviously, this is the same as
showing the distribution of $M_{\rm crit}/M_{_{\rm TOV}}$ for $j=j_{\rm
  Kep}$ as done in Fig.~\ref{fig:BR_Mtov2p2}. As can be readily observed,
the median of our results is surprisingly robust even when only $10^{2}$
EOSs are used and the corresponding median values is well within
$1$-$\sigma$ away from the result obtained with $10^{6}$ EOSs. This
notwithstanding, the overall behaviour of the PDF can be seen to be
significantly more noisy for the lower statistics cases, and, in
particular, far less EOSs sample the tails of the distribution. For this
reason, it is still important to use large statistics so that the $100\%$
confidence interval can be trusted to span the whole physically allowed
region for $\mathcal{R}$.

Next, we discuss how to obtain a new upper limit of the maximum mass of
nonrotating stars after using the newly estimate for $\mathcal{R}$ and
the assumption that the GW170817 remnant collapsed while still rotating
at a frequency close to the mass-shedding limit. Under these assumptions,
it is possible to obtain a first rough upper bound on $M_{_{\rm TOV}}$ by
simply observing that $M_{\rm max} < M_{\rm GW170817} =
2.74^{+0.04}_{-0.01}\,M_{\odot}$, and by subsequently using the value of
$\mathcal{R}:=M_{\rm max}/M_{\rm TOV}$ obtained by the quasi-universal
relation presented in the main text. This estimate can be further refined
by making use of the iterative procedure outlined below.

\begin{enumerate}
\item We construct an EOS ensemble as discussed in the main text
  recalling that the dependence of $\mathcal{R}$ on the lower bound
  imposed on $M_{\rm TOV}$ is weak but non-negligible (we here set of
  $M^{(-)}_{\rm TOV}=2.08\,M_{\odot}$).
\item We determine the quasi-universal ratio $\mathcal{R}^{n=1}$, where
  $n$ is the iteration number.
\item Assuming that GW170817 collapsed to a black hole, we determine the
  upper bound $M^{(+)}_{\rm TOV}$.
\item We construct a new ensemble of EOSs, but this time imposing as
  upper bound on $M_{\rm TOV}$ the value determined in point 3.
\item We recompute $\mathcal{R}^{n}$ and keep iterating from point
  3. until convergence has been reached. We find that $n=3$ is sufficient
  to obtain differences below $1\%$ between $\mathcal{R}^{n}$ and
  $\mathcal{R}^{n+1}$.
\end{enumerate}
Following this iterative procedure we have derived the upper limit for
the maximum mass of nonrotating neutron stars reported in the main text:
$M^{(+)}_{\rm TOV} = 2.24^{+ 0.07}_{-0.11}$.

\subsection*{Quasi-universal relations for rapidly rotating stars}
\label{sec:qurrrs}

Following~\citet{Pappas2014}, we define the dimensionless moment of
inertia and quadrupole respectively as $\bar{I}:=I/M^3$ and
$\bar{Q}:=Q/M^3/j^2$, and seek a quasi-universal expression for $\bar{I}$
as a function of $j$ and $\bar{Q}$ via the ansatz
\begin{equation}\label{eq:ILQfit}
  \sqrt{\bar{I}} = A_1 + A_2\left( \sqrt{\bar{Q}} - \xi_0 \right) +
  A_3\left( \sqrt{\bar{Q}} - \xi_0 \right)^2\,,
\end{equation}
where
\begin{align} 
  A_2 := B_1 + B_2~j + B_3~j^2\,, \quad {\rm and} \quad A_3 := C_1 + C_2~j +
  C_3~j^2\,,
  \label{eq:A3}
\end{align}
with $A_1$, $B_1-B_3$, $C_1-C_3$, $\xi_0$ fitting parameters. The
corresponding values for the EOS ensembles with $M^{(-)}_{\rm
  TOV}=2.2~M_{\odot}$ and when the pQCD constraints are taken into
account or not are reported in Tab.~\ref{tab:ILQ}.

\begin{table}
  \setlength{\tabcolsep}{2.2pt} 
  \renewcommand{\arraystretch}{1.0} 
  \center
  \caption{Best-fit coefficients for the $I$-$Q$ quasi-universal relation
    proposed by \citet{Pappas2014} [\cf Eq.~\eqref{eq:ILQfit}] with
    $M^{(-)}_{\rm TOV}=2.2~M_{\odot}$.}
 \label{tab:ILQ}
 \begin{tabular}{ccccccccc} 
   \hline
   pQCD & $A_1$ & $B_1$ & $B_2$ & $B_3$ & $C_1$ & $C_2$ & $C_3$ & $\xi_0$ \\ 
   \hline
   yes &  $2.07$ & $0.764$ & $0.238$ & $0.816$ & $0.143$  & $0.0316$ & $-0.128$ & $0.908$ \\ 
   no  &  $2.12$ & $0.705$ & $0.537$ & $0.519$ & $0.175$  & $-0.100$ & $0.014$  & $0.951$ \\ 
   \hline 
 \end{tabular}
\end{table}

\end{document}